\documentclass[aps,prb,twocolumn,superscriptaddress]{revtex4}

\usepackage{graphicx}

\newlength{\figwidth}

\divide\figwidth by 2

\begin{document}


\title{AgCuVO$_4$: a quasi one-dimensional {\it S} = 1/2 chain compound}

\author{A. M\"oller}
\affiliation{Institut f\"ur Anorganische Chemie, Universit\"at zu
K\"oln, Greinstr.~6,  50939 K\"oln, Germany}\thanks{current address:
Department of Chemistry and TCSUH, University of Houston, Houston,
Texas 77204-5003, USA}

\author{M. Schmitt} \author{W. Schnelle} \author{T. F\"orster}
\author{H. Rosner} \affiliation{Max Planck Institute for Chemical
  Physics of Solids, N\"othnitzer Str.~40, 01187 Dresden, Germany }



\date{\today}
\begin{abstract}
We present a joint experimental and computational study of the
recently synthesized spin 1/2 system silver-copper-orthovanadate
AgCuVO$_4$ [A. M\"oller and J. Jainski, Z. Anorg. Allg. Chem.
\textbf{634}, 1669 (2008)] exhibiting chains of $trans$
corner-sharing [CuO$_4$] plaquettes. The static magnetic
susceptibility and specific heat measurements of AgCuVO$_4$ can be
described to a good approximation by the Bonner-Fisher spin-chain
model with $J_{\mathrm{intra}} \approx$ 330 K.  Evidence for a
N\'eel type of order at $\sim$ 2.5\,K is obtained from the specific
heat and corroborated by ESR studies. To independently obtain a
microscopically based magnetic model, density functional electronic
structure calculations were performed. In good agreement with the
experimental data, we find pronounced one-dimensional magnetic
exchange along the corner-sharing chains with small inter-chain
couplings. The difference between the experimentally observed and
the calculated ordering temperature can be assigned to a sizable
inter-chain frustration derived from the calculations.
\end{abstract}


\maketitle

\section{Introduction}

The quest for materials that form quasi one-dimensional magnetic
subunits has evolved into an important part of modern solid-state
physics and chemistry over the last one or two
decades.\cite{dagotto} Concerning quasi one-dimensional spin chains,
the fundamentally different behavior of systems of even- and odd
spin numbers has been one of the most spectacular theoretical
findings by Haldane already in 1983.\cite{haldane} In subsequent
years it was realized that a large variety of exotic ground states
and different unconventional elementary excitations are realized in
low-dimensional quantum {\it S} = 1/2 systems, in many cases due to
frustration effects. In particular, the ground state properties of a
{\it S} =1/2 chain with nearest and next-nearest neighbor
interactions were calculated by Bursill \emph{et al.}\cite{Bursill}
with the result of a spiral spin-order, depending sensitively on the
ratio of the interaction parameters $J_{ij}$ of the corresponding
Heisenberg Hamiltonian

\begin{equation}
\hat{H} =  \sum_{<ij>} J_{ij} \hat{S}_i \hat{S}_{j}. \label{ham}
\end{equation}

On the experimental side, the field of low-dimensional quantum
magnetism has recently seen a considerable impetus from another
subject of research activity, namely
multiferroicity.\cite{schrettle} Naito \emph{et al.}\cite{naito}
found that the spiral spin order of {\it S} = 1/2 chains in
LiCuVO$_4$ is able to induce a ferroelectric polarization at
temperatures below $\approx$ 2.4 K. A similar behavior has been
observed for LiCu$_2$O$_2$.\cite{Park} The appropriate magnetic
model \cite{masuda04,gippius04,drechsler,masuda05} and the
theoretical description of this phenomenon has been controversially
discussed in the literature recently, from both a phenomenological
\cite{mostovoy} and microscopical point of
view.\cite{katsura,sergienko}

In order to gain further insight into the phenomena and mechanisms
at work, it is important to study similar systems with comparable
characteristics. From a structural point {\it S} = 1/2 chains of
interlinked [CuO$_4$] plaquettes may connect via edges or corners.
Whereas the latter systems show typically antiferromagnetic nearest
neighbor (NN) couplings, the edge-shared systems (i.e. LiCuVO$_4$,
LiCu$_2$O$_2$) exhibit either ferro- or antiferromagnetic NN
couplings according to their Cu-O-Cu bond angles close to 90$^\circ$
in agreement with the Goodenough-Kanamori-Anderson
rules.\cite{goodenough} A further structural feature is whether the
[CuO$_4$] plaquettes of the chain are all orientated in a co-planar
or corrugated/buckled fashion. This will significantly affect the
intra-chain exchange parameters $J_{\mathrm{intra}}$ as well.  For
example, for the (planar) chain system Sr$_2$CuO$_3$ of
corner-sharing [CuO$_4$] plaquettes one finds a pronounced
contribution $J_{\mathrm{NNN}}$ from next-nearest neighbors (NNN)
with $J_{\mathrm{NN}}/J_{\mathrm{NNN}}\approx$ 25.\cite{rosner97}

Furthermore, the influence of inter-chain interactions needs
to be considered and evaluated especially in the very low temperature
regime. Thereby, a classification of the one-dimensional systems
regarding their magnetic properties may be achieved, e.g.
Spin-Peierls transition (dimerised model) or magnetically ordered
exhibiting either classical N\'eel or spiral spin order,
respectively.

\begin{figure}
\begin{center}
\resizebox{0.8\figwidth}{!}{
\includegraphics*[angle=0]{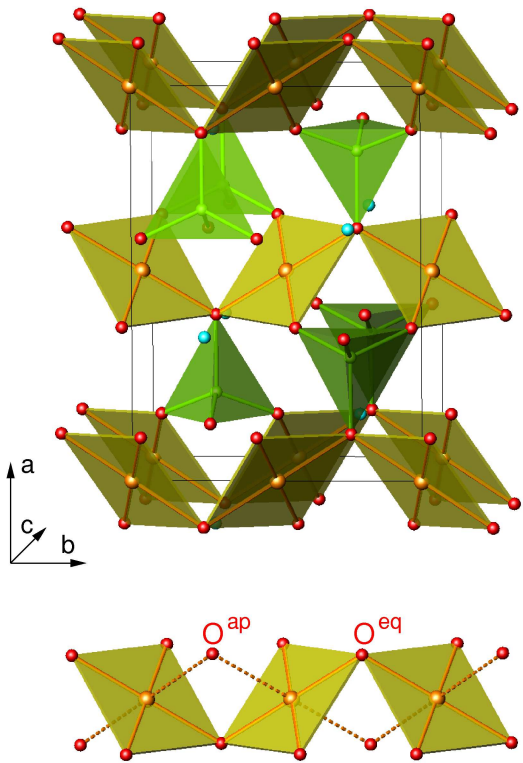}}
\caption{(Color online) Top: Crystal structure of AgCuVO$_4$.
Green: [VO$_4$] tetrahedra. Yellow: [CuO$_4$] distorted
square-planar entities. Light blue: Ag atoms. Bottom: Chain-like connectivity
of the [CuO$_4$] units in AgCuVO$_4$. (Cu-O$^{\mathrm{eq}}$(bold lines),
Cu-O$^{\mathrm{ap}}$ (dotted lines))}\label{ez}
\end{center}
\end{figure}

In this paper, we present a combined experimental and theoretical
study of the {\it S} = 1/2 chain compound AgCuVO$_4$
(Fig.~\ref{ez}), which was synthesized recently.\cite{judith} A
classification of this system only based on its crystal structure is
far from obvious: Whereas the [CuO$_4$] plaquettes in AgCuVO$_4$
form corner-shared chains, the Cu-O-Cu bond angles are rather close
to the typical angles in edge-shared cuprate chains. Our combined
approach shows consistently that AgCuVO$_4$ can be described as a
quasi one-dimensional {\it S} = 1/2 NN-only Heisenberg chain in very
good approximation.

\section{Methods}

Powder samples of AgCuVO$_4$ were synthesized by solid state
reactions of CuO and $\alpha$-AgVO$_3$ (obtained from Ag$_2$O and
V$_2$O$_5$) at 700\,K in alumina crucibles in air. Powder pellets
were pressed using a self-made molding press and sintered for
several days at 700\,K. These samples were then used for
measurements of the physical properties. The phase purity of all
samples was checked by powder X-ray diffraction. Low temperature
X-ray diffraction patterns were measured on a Stoe \& Cie.
diffractometer equipped with a closed-cycle Helium cryostat (Lake
Shore) using Cu-K$_{\alpha}$ radiation between 20 and 293\,K. All
measured diffraction patterns could be indexed and refined in the
space group \emph{Pnma} discussed below, revealing no structural
transition in the investigated range.

The static magnetic susceptibility of AgCuVO$_4$ was measured in
external magnetic fields of 0.1\,T and 0.5\,T in field-cooled (fc)
and zero-field-cooled (zfc) mode (SQUID magnetometer; MPMS, Quantum Design).

The electron spin resonance (ESR) measurements were performed at
X-band frequencies ({\it f} = 9.4~GHz) using a standard spectrometer
together with a He-flow cryostat that allows to vary the
temperature from 2.6 to 300\,K. ESR probes the absorbed power $P$ of a
transversal magnetic microwave field as a function of a static and
external magnetic field $B$. To improve the signal-to-noise ratio, we
used a lock-in technique by modulating the static field, which yields
the derivative of the resonance signal $dP/dB$.

The heat capacity was determined on the same sample by a
relaxation-type method ($m \approx 11$\,mg; PPMS, Quantum Design) in
external magnetic fields of 0, 3, 6, and 9\,T.

For the electronic structure calculations the full-potential
local-orbital scheme FPLO (version: fplo7.00-28) within the local
(spin) density approximation (L(S)DA) was used.\cite{fplo1} In the
scalar relativistic calculations the exchange and correlation
potential of Perdew and Wang was chosen.\cite{PW92} To consider the
strong electron correlations for the Cu$^{2+}$ (3$d^9$)
configuration, we use the LSDA+$U$~\cite{fplo2} approximation
varying $U_{d}$ in the physically relevant range from 6 -- 8.5\,eV.
The LDA results were mapped onto an effective tight-binding model
(TB) and subsequently to a Hubbard and a Heisenberg model.

\section{Crystal structure}
Fig.~\ref{ez} shows a perspective view of the crystal structure of
AgCuVO$_4$. The compound crystalizes in the orthorhombic space group
\emph{Pnma} with lattice parameters $a$ = 9.255(1)~\AA, $b$ =
6.778(1)~\AA~ and $c$ = 5.401(1)~\AA, as determined by single
crystal X-ray diffraction at room temperature.\cite{judith} Isolated
[VO$_4$] tetrahedra (containing 'nonmagnetic' V$^{5+}$) are
connected to the Cu$^{2+}$ ions as bridging complex ions within the
chain as well as between neighboring chains of $trans$
corner-sharing [CuO$_4$] plaquette entities along the $b$ axis.
These chains are furthermore separated from each other by isolated,
nonmagnetic Ag$^+$ ions, also shown in Fig.~\ref{ez}. A more
detailed description of the crystal structure is given by M\"oller
\emph{et al.}\cite{judith}

The magnetic Cu$^{2+}$ ions ({\it S} = 1/2) are coordinated by four
oxygen atoms in a square-planar fashion with a typical average
Cu-O$^{\text{eq}}$ distance of 1.995~\AA, whereas the distance to
the two apical oxygen atoms (O$^{\text{ap}}$), designated by dotted
lines in Fig.~\ref{ez}, is 2.511(4)~\AA. Considering the elongated
octahedral coordination of Cu$^{2+}$ a $d_{x^2-y^2}$ ground state
character can be assumed. The bridging angle
$\angle$(Cu-O$^{\text{eq}}$-Cu) within the chains is
113.0(2)$^{\circ}$, indicative of predominantly antiferromagnetic
interactions of the spins according to the
Goodenough-Kanamori-Anderson rules.\cite{goodenough} Although
Fig.~\ref{ez} suggests an edge-sharing connectivity within the
chains, the effective exchange via the apical oxygen atom
(O$^{\text{ap}}$) is almost negligible, since the related $d_{z^2}$
is fully occupied. Any superexchange via O$^{\text{ap}}$ would be
expected to give a ferromagnetic coupling in relation to the
bridging angle (Cu-O$^{\text{ap}}$-Cu) of 85$^{\circ}$. Therefore,
one might consider this system as a quasi one-dimensional {\it S} =
1/2 antiferromagnet build of $trans$ corner-sharing [CuO$_4$]
plaquettes. However, it should be noted that AgCuVO$_4$ is not
isotypic with LiCuVO$_4$ which contains edge-sharing [CuO$_4$]
plaquette entities. Thus, distinct differences with respect to the
magnetic interaction pathways associated with the Cu$^{2+}$
$d_{x^2-y^2}$ ground state occur.


\section{Results and discussion}

\subsection{Magnetic susceptibility}

\begin{figure}
\begin{center}
\resizebox{0.98\figwidth}{!}{
\includegraphics*[angle=0]{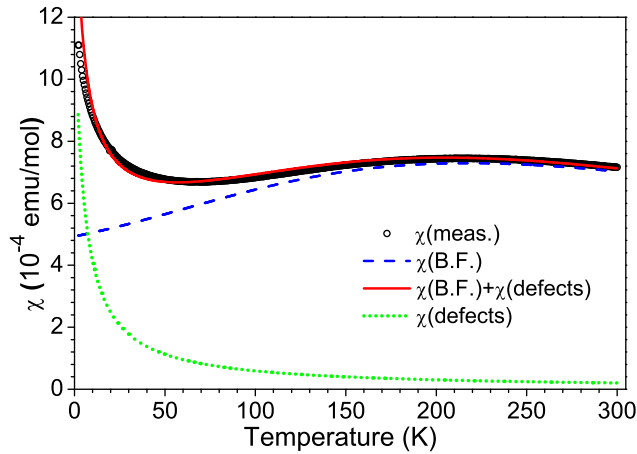}}
\caption{(Color online) Temperature dependence of the static
magnetic susceptibility of AgCuVO$_4$ (0.1\,T, fc), black circles.
$\chi$(B.F.) gives the result for an isotropic chain (Bonner-Fisher
model, dashed blue line) and $\chi$(defects) present the impurity
contribution, dotted green line. The total fit is given by the red
line.}\label{chi}
\end{center}
\end{figure}

The temperature dependence of the static magnetic susceptibility
measured in a field of 0.1 T in fc mode is plotted in
Fig.~\ref{chi}. The broad maximum in $\chi(T)$ at around 200\,K is a
typical feature of a low-dimensional magnetic system. Towards lower
temperatures, below approximately 50\,K, a sharp upturn in the form
of a 'Curie-tail' occurs, which most probably originates from
paramagnetic impurities. Measurements at 0.1\,T and 0.5\,T showed
identical behavior in fc and zfc mode (not shown here). At higher
temperatures, the magnetic susceptibility $\chi(T)$ can be described
according to the isotropic one-dimensional {\it S} = 1/2 model of
Bonner and Fisher \cite{bonner} using the spin-spin Hamiltonian in
the form of $\hat{H} = J_{\mathrm{intra}} {\sum} \hat{S}_i
\hat{S}_{1+i}$. A magnetic coupling constant of
$J^{\chi}_{\mathrm{intra}} \approx$ 335\,K is derived for
AgCuVO$_4$, see dashed blue line in Fig.~\ref{chi}. We used a
typical constant value for the $g$-factor ($g_{\text{average}}
\approx 2.10$) for the determination of $J^{\chi}_{\mathrm{intra}}$
within the isotropic Heisenberg model. At low temperatures the
magnetic contribution of defects becomes evident. Including 1.5 \%
paramagnetic {\it S} = 1/2 impurities (defects, dotted green line)
with a Weiss temperature of $\Theta_{\mathrm{imp}} \approx$ -5\,K we
obtain a fit in good agreement with the experimental data for the
temperature range from 10 to 300\,K (red line of Fig.~\ref{chi}).

The effective antiferromagnetic coupling indicates a dominant
$d_{x^2-y^2}$ ground state character of the Cu$^{2+}$ ion in
agreement with a superexchange path via the bridging oxygen atoms
O$^{\text{eq}}$ with an angle of 113$^{\circ}$ in this case. Of
course, this highly oversimplified picture relating to structural
arguments calls for a more detailed analysis based on specific heat,
ESR measurements and LDA calculations.

\subsection{Electron spin resonance}

\begin{figure}
\begin{center}
\resizebox{0.90\figwidth}{!}{
\includegraphics*[angle=0]{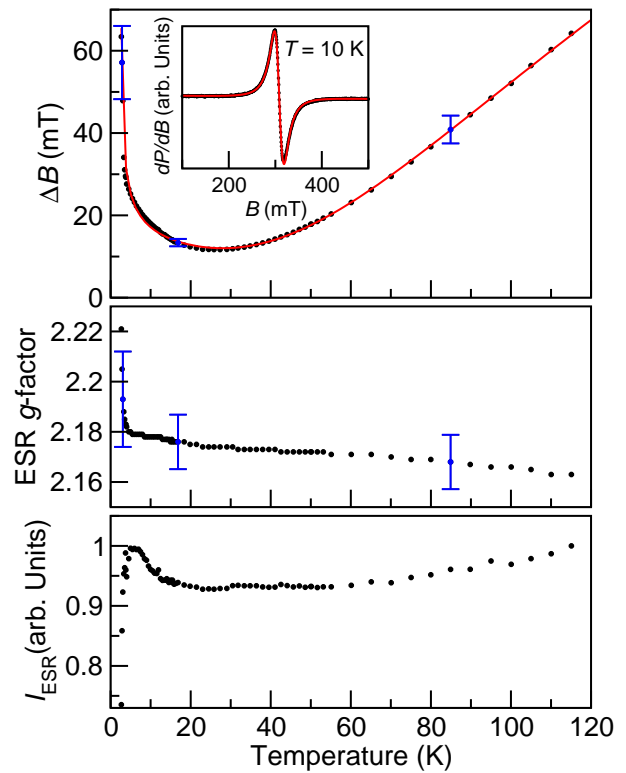}}
\caption{(Color online) The inset shows a typical ESR signal (black
circles) with a Lorentzian fit (red line) given. The temperature
dependence of the signals line width (black circles) and a fit with
Eq. \ref{dBFitform} to the data (red line) is shown for AgCuVO$_4$
in the top main panel. The two lower panels show the temperature
dependence of the $g$-factor (middle) and the intensity of the ESR
signal (bottom panel) for AgCuVO$_4$.}\label{esr}
\end{center}
\end{figure}

Fig.~\ref{esr} shows a typical ESR signal at $T$ = 10\,K (inset) and
the temperature behavior of the line width (top panel) for
AgCuVO$_4$. At temperatures $T >$ 4\,K we observe a well defined
signal which could be nicely fitted with a single Lorentzian line
(red line in inset) providing the ESR parameters line width $\Delta
B$ and $g$-factor ($g=h\nu/\mu_BB_{\mathrm{res}}, B_{\mathrm{res}}$-
resonance field). Below 4\,K an additional signal with a twice
larger line width and nearly the same resonance field occurs. This
second signal might be related to the anisotropy of the ESR
parameters, but could also originate from impurities or indicate a
resonance mode resulting form an antiferromagnetic ordered state.

In the two lower panels in Fig.~\ref{esr} the temperature behavior
of the ESR intensity and the measured $g$-factor is shown. The
observed $g$-values of $\approx$ 2.17(1) are typical for Cu$^{2+}$
in a distorted octahedral environment and in fair agreement with the
static susceptibility fitted by the Bonner-Fischer model. Above
10\,K the intensity follows the magnetic susceptibility $\chi(T)$
(see Fig.~\ref{chi}) reproducing the broad minimum around 40\,K.
Below $T$ = 4\,K the ESR intensity reduces rapidly towards lower
temperatures indicating a magnetic ordering below 4\,K. A similar
behavior above $T_N$ is observed in the spin chain compound
CuSiO$_3$.\cite{sichelschmidt02}

The divergence of the $g$-factor and the line width at low
temperatures also indicate a magnetic phase transition below 4\,K.
The temperature dependence of the line width $\Delta B(T)$ is
comparable with that of LiCuVO$_4$\cite{nidda} and
CuGeO$_3$\cite{eremina03} showing a pronounced minimum between 10
and 40\,K. We fitted $\Delta B(T)$ with a classical critical
divergence at $T_N$ (first term in Eq. (\ref{dBFitform}) below) and
an empirical expression for the high temperature
part.\cite{eremina03}(second term in Eq.~(\ref{dBFitform})):

\begin{equation}
\Delta B=C\left(\frac{T}{T_N}-1\right)^{-p}+\Delta B(\infty)
\exp{\left(- \frac{C_1}{T+C_2}\right)} \label{dBFitform}
\end{equation}

$C$, $p$, $T_N$, $\Delta B(\infty)$, $C_1$, and $C_2$ are treated as
6 fitting parameters. Table \ref{Fitresult} summarizes the fit
results.

\begin{table}[h]
\caption{\label{Fitresult} Fit parameters according to
equation~(\ref{dBFitform}) for the line width of the ESR signals for
AgCuVO$_4$}
\begin{ruledtabular}
\begin{tabular}{c   c   c  c   c   c  c}
 $C$ (mT)& $p$ &$T_N$ (K) & $\Delta B(\infty)$ (mT)& $C_1$ (K) & $C_2$ (K)\\
\hline 35$\pm$0.7 & 0.35$\pm$0.05 &2.5$\pm$0.2 & 294$\pm$50 &
199$\pm$30 & 6$\pm$5
\end{tabular}
\end{ruledtabular}
\end{table}

The fit results on the first term in Eq.~(\ref{dBFitform}) depend
only weakly on the parameters of the second term and vice versa.
This holds especially for $T_N$. The influence of the other
parameters is included in the uncertainty displayed in Table
\ref{Fitresult}.

$C_1$ is related to the order of magnitude of the isotropic exchange
constant, because the parameter indicates the transition from the
strongly correlated one-dimensional regime at low temperatures $T
\ll J_{\mathrm{intra}}$ to the purely paramagnetic regime $T \gg J_{\mathrm{intra}}$.
$C_2$ indicates the influence of the low-temperature phase
transition (long-range magnetically ordered phase) on the line
broadening. It is necessary to recall that this purely empirical
parametrization has no underlying microscopic picture. Nevertheless,
the ESR corroborates the antiferromagnetic ground
state and the onset of long-range order setting in at $\approx$ 2.5
K (see section on the specific heat capacity below).

\subsection{Specific heat capacity}

\begin{figure}
\begin{center}
\resizebox{0.95\figwidth}{!}{
\includegraphics*[angle=0]{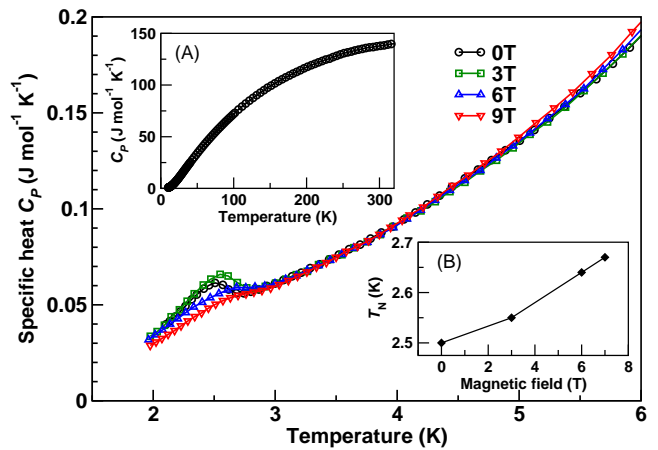}}
\caption{(Color online) Temperature dependence of the specific heat
of AgCuVO$_4$ for different magnetic fields. Inset (A): Large
temperature range in zero field. Inset (B): Field dependence of the
N\'eel temperature.}\label{cp}
\end{center}
\end{figure}

\begin{figure}
\begin{center}
\resizebox{0.98\figwidth}{!}{
\includegraphics*[angle=0]{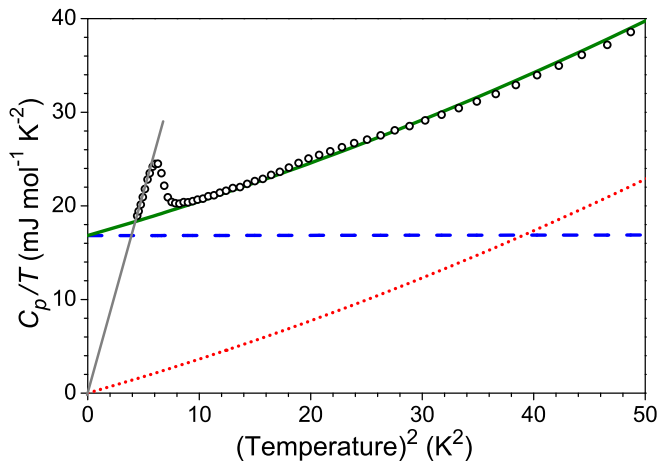}}
\caption{(Color online) Magnetic heat capacity $C_P/T$ at zero field
versus $T^2$ (black circles) for AgCuVO$_4$ including the fit to
equation (\ref{lattice}) (green line) as well as $C_m/T$ (dashed
blue line) and $C_L/T$ (dotted red line). The grey line presents a
linear fit below $T_N$, see text.}\label{cpfit}
\end{center}
\end{figure}

Inset A of Fig.~\ref{cp} shows the measured specific heat capacity,
$C_p(T)$, in zero field for AgCuVO$_4$ below 320\,K. The field
dependence of $C_p(T)$ is given in the main panel of Fig.~\ref{cp}.
Below 3\,K an anomaly is clearly visible in lower fields, which
broadens and smears out with increasing field. The field dependence
indicates that this effect can be attributed to an
antiferromagnetically ordered ground state. The shift of the the
N\'eel temperature with increasing field from 2.5\,K to almost
2.7\,K is shown in the inset (B) of Fig.~\ref{cp}. We have fitted
$C_p(T)/T$ versus $T^2$ below 10\,K (Fig.~\ref{cpfit}) to
Eq.~(\ref{lattice}) in order to obtain the coefficients ($\gamma$
and $\beta_1$) related to the magnetic ($C_m(T)$) and lattice
($C_l(T)$; harmonic lattice approximation) part of the specific
heat.

\begin{equation}
C_p(T)/T = \gamma + \beta_1T^2 + \beta_2T^4 \label{lattice}
\end{equation}

The lattice contribution (phonons) is associated with $\beta_1$ =
0.34(5) mJ mol$^{-1}$ K$^{-4}$ and allows an estimate of the Debye
temperature $\Theta_{D} = (12 \pi n N_A k_B/5\beta_1)^{1/3} \approx$
342\,K, a typical value for $^1_{\infty}$[CuO$_{4/2}$] compounds
with edge-sharing connectivity.\cite{sl_walter} The value of
$\beta_2$ = 0.00234(4) mJ mol$^{-1}$ K$^{-6}$ is of the expected
magnitude ($\l$ 1\% $\beta_1$). The magnetic part of the specific
heat is presented by $\gamma$ = $C_m/T$. $C_m$($T$)/$T$ is found to
be almost constant at low temperatures for an isotropic {\it S} =
1/2 chain and related to the magnetic exchange parameter,
$J_{\mathrm{intra}}$ =
2$N_Ak_B$/(3$C_p^{(T\rightarrow0)}/T$).\cite{klumper} With $\gamma$
= 16.9(1) mJ mol$^{-1}$ K$^{-2}$ we find within the Heisenberg model
$J^{C_p}_{\mathrm{intra}}$ of $\approx$ 330 $\pm$ 10\,K in excellent
agreement with the value derived from the fit to the susceptibility
data.

The long-range order setting in below 3\,K occurs from inter-chain
couplings and the N\'eel temperature, $T_N$, associated with this
transition can be used to estimate the exchange parameter
$J^{C_p}_{\mathrm{inter}}$ of $\approx$ 0.75 K from Eq. (\ref{neel})
\cite{schulz}
\begin{equation}
J_{\mathrm{inter}} = T_N/(1.28\sqrt{ln(5.8J_{\mathrm{intra}}/T_N}) \label{neel}
\end{equation}

In principle the same result is obtained from a more recent
theoretical approach \cite{Irkhin} to estimate the inter-chain
coupling from $T_N$. Here, the intra-chain coupling parameters
$J_{\mathrm{intra}}$ derived from $C_p(T)$ and $\chi(T)$ data are
not only consistent but almost identical. Furthermore,
$J_{\mathrm{intra}}$ is found to be $\approx$ 440 times larger than
$J_{\mathrm{inter}}$. Therefore, AgCuVO$_4$ can be regarded as a
quasi one-dimensional {\it S} = 1/2 system at $T \gg$ 3\,K. For a
more detailed analysis and evaluation of $J_{\mathrm{inter}}$ in
this case see section on the electronic structure below.

In Fig.~\ref{cpfit} a linear fit below $T_N$ ($T^2 <$ 5.5\,K) is
included. The extrapolated intersection with the origin is in
agreement with the insulating properties of AgCuVO$_4$ and reveals the
magnon contribution to the total specific heat capacity, $C_p$, which
follows approximately a $T^3$ law typical for an antiferromagnet.

\subsection{Electronic structure}
To gain microscopic insight into the electronic and magnetic
properties of the system we carried out band structure calculations
based on density functional theory (DFT) and subsequent model
calculations.

\begin{figure}
\begin{center}
\resizebox{0.99\figwidth}{!}{
\includegraphics*[angle=0]{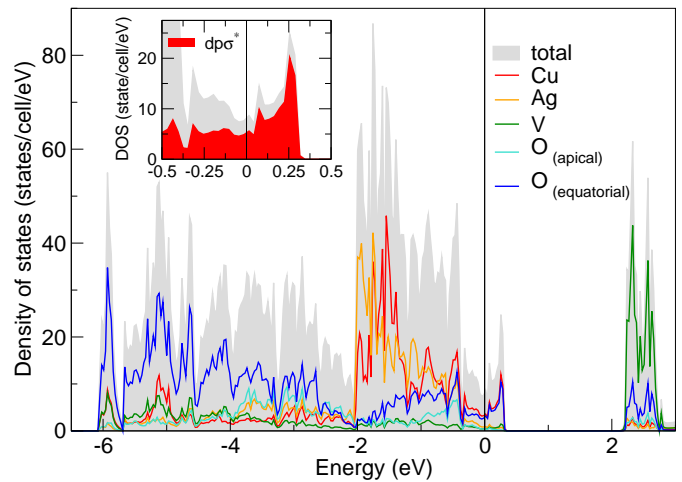}}
\caption{(Color online) Calculated total and atom resolved density
of states for AgCuVO$_4$. The Fermi level is at zero energy. Inset:
Orbital-resolved density of states for the $dp\sigma^*$ states of
the antibonding half-filled band.}\label{dos}
\end{center}
\end{figure}

In Fig.~\ref{dos} the obtained density of states (DOS) of the
valence states with a band width of about 6\,eV is shown. This band
width is rather typical compared with other cuprates revealing a
chain-type of structural feature like CuGeO$_3$ \cite{rosner2000} or
Sr$_2$CuO$_3$ \cite{rosner97}. The valence band is dominated by Ag,
Cu and O states. The quite narrow Ag 4$d$ contribution between
$-$2\,eV and $-$0.5\,eV indicates a Ag$^{+}$ cation. For Cu and V
the calculations yield magnetic Cu$^{2+}$ and non-magnetic V$^{5+}$
as could be expected from the crystal structure in terms of their
coordination spheres (see Fig.~\ref{ez}): Cu and O form strongly
distorted [CuO$_6$] octrahedra with considerably shorter
Cu-O$^{\mathrm{eq}}$ bonds in the equatorial plane leading to the
characteristic [CuO$_4$] plaquettes, whereas the non-magnetic
V$^{5+}$ is tetrahedrally coordinated, [VO$_4$]. Magnetically active
V$^{4+}$ usually appears in drastically distorted coordination
spheres (e.g. square-pyramidal).

\begin{figure}
\begin{center}
\resizebox{0.99\figwidth}{!}{
\includegraphics*[angle=0]{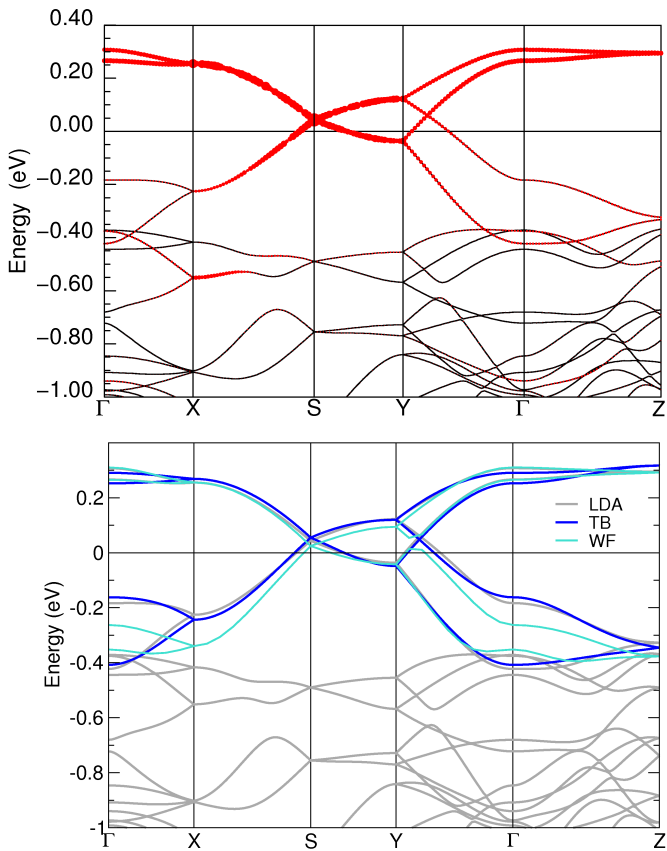}}
\caption{(Color online) Antibonding $dp\sigma^*$ bands from the LDA
calculations together with the band from a direct tight binding (TB)
fit and from a Wannier function (WF) based mapping.}\label{tbband}
\end{center}
\end{figure}

As a consequence of the Cu$^{2+}$ state the anti-bonding Cu-O
dp$\sigma^*$ states of the system are half filled. Corresponding to
the 4 Cu per unit cell, four half-filled bands cross the Fermi energy
$\varepsilon_F$ (see Fig.~\ref{tbband}) in our LDA calculation. This
metallic solution is in contradiction to the insulating character of
the compound concluded from the red color of the crystals and a well
known shortcoming of the LDA calculations. As LDA underestimates the
strong correlations of the Cu$^{2+}$ ($3d^9$) configuration, these
missing correlations can be taken into account by mapping the
relevant LDA bands onto a TB model and subsequently onto an Hubbard
and Heisenberg model. Alternatively the strong Coulomb repulsion in
the Cu$3d$ shell can be considered explicitly in a mean-field like
approach using the LSDA+$U$ scheme.

In many cuprates the anti-bonding dp$\sigma^*$ states are well
separated from the lower-lying valence states, while in AgCuVO$_4$
these states overlap with the lower lying part of the valence due to
a sizable admixture of other orbitals.  The inset of Fig.~\ref{dos}
shows the orbital-resolved DOS of the Cu-O dp$\sigma^*$ states in
comparison with the total DOS. In the region between $\varepsilon_F$
and 0.25\,eV the dp$\sigma^*$ states clearly dominate the
antibonding bands, whereas from $\varepsilon_F$ to $-$0.5\,eV
contributions from other orbitals increases notably. Accordingly,
the hybridization with lower lying parts of the valence band is also
visible in the upper panel of Fig.~\ref{tbband} where the band
characters of the dp$\sigma^*$ states spread out to lower energies
(mainly around $\Gamma$ and $X$).

\begin{figure}
\begin{center}
\resizebox{0.85\figwidth}{!}{
\includegraphics*[angle=0]{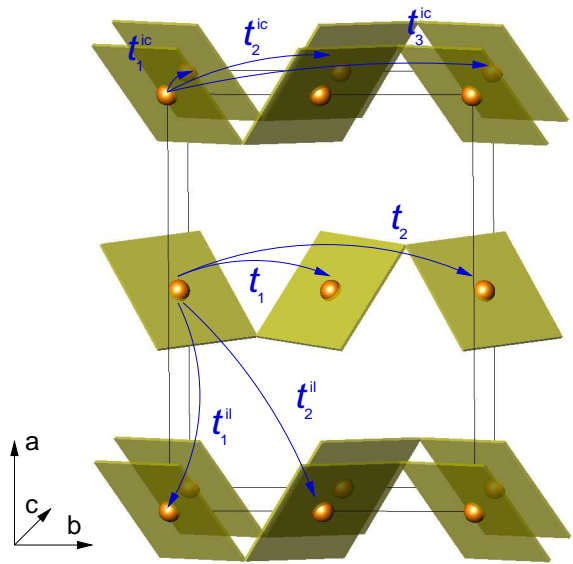}}
\caption{(Color online) Sketch of the TB model for
AgCuVO$_4$}\label{tbmodel}
\end{center}
\end{figure}

\begin{figure}
\begin{center}
\resizebox{0.85\figwidth}{!}{
\includegraphics*[angle=0]{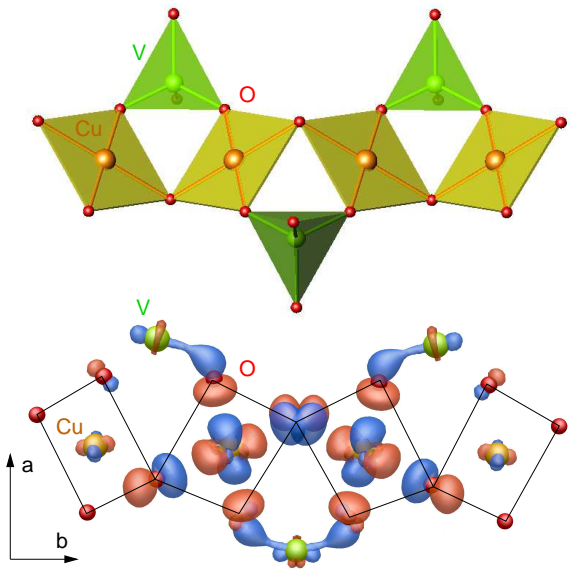}}
\caption{(Color online) Cu 3$d_{x^2-y^2}$ related Wannier functions
for a chain segment in AgCuVO$_4$.}\label{wannier}
\end{center}
\end{figure}

The sizable admixture of other valence states to the antibonding
bands that are responsible for the magnetic interactions in the
system impedes a straight forward mapping to an effective one-band
TB model using a least square fit procedure. The ambiguousness in
the selection of the relevant bands, especially between $\Gamma$ and
$X$ and $\Gamma$ and $Z$, respectively, can be removed applying the
Wannier function technique (see Fig.~\ref{wannier}). The resulting
leading transfer terms for both approaches are sketched in
Fig.~\ref{tbmodel} and the values are given in Table~\ref{tandJ}.
The corresponding bands are highlighted in Fig.~\ref{tbband} (lower
panel) on top of the LDA band structure. It can be clearly seen that
in the upper part the least square TB fit and the Wannier function
derived bands nearly coincide, while for the lower lying region with
stronger admixture (see inset Fig.~\ref{dos}) both approaches show
sizable deviations. This is mostly reflected in the leading nearest
neighbor (NN) transfer integral $t_1$, whereas the much smaller
coupling to further neighbors are mostly unaffected. On the other
hand, the good agreement between both methods justifies the
application of an effective one-band picture.

\begin{table}[hbt]
\caption{\label{tandJ} Leading coupling terms for AgCuVO$_4$ from TB
and WF approach.}
\begin{ruledtabular}
\begin{tabular}{c|c c c c c c c}
$t_i$/meV & $t_1$ & $t_2$  & $t_1^{\mathrm{ic}}$ & $t_2^{\mathrm{ic}}$ & $t_3^{\mathrm{ic}}$& $t_1^{\mathrm{il}}$& $t_2^{\mathrm{il}}$ \\
\hline
$t_i$(TB)& -145 & 2 & 17 & 7 & 6/-19 & -16 & -11/0\\
$t_i$(WF)& -157 & 0 & 18 & 6 & 5/-20 & -13 & -10/3\\
\end{tabular}
\end{ruledtabular}
\end{table}

From our mapping, we obtain a clear physical picture for the
relevant interactions in the system: We find weakly interacting
chains along the $b$ direction with essentially only NN coupling
$t_1$ (see Table~\ref{tandJ}). Consistent with the Cu-O-Cu bond
angle of about 113$^\circ$ this leading transfer is closer to the
values for edge-shared chain geometry (e.g.
CuGeO$_3$\cite{rosner2000}, Cu-O-Cu bond anlge of about 99$^\circ$,
$t_1$~$\approx$~175\,meV) than for the couplings of the
corner-shared chains (e.g. Sr$_2$CuO$_3$\cite{rosner97}, Cu-O-Cu
bond angle of 180$^\circ$, $t_1$~$\approx$~410\,meV). This vicinity
to the edge-shared chain compounds raises the question of the
relevance of ferromagnetic contributions to the NN exchange $J_1$.
Mapping the TB model via a Hubbard to a Heisenberg model (in the
limit of strong correlations and at half filling) to describe the
low lying magnetic excitations only yields the antiferromagnetic
parts $J_i^{\mathrm{AF}} = 4t_i^2/U_{\mathrm{eff}}$ of the total
exchanges $J_i$. The ferromagnetic contributions can be estimated
comparing the TB derived exchange $J_i^{\mathrm{AF}}$ with the
result of LSDA+$U$ calculations for magnetic super cells. Using a
standard one-band value $U_{\mathrm{eff}}$~=~4\,eV\cite{rosner2000}
we obtain for the NN exchange $J_1^{\mathrm{AF}}$~=~23~$\pm$~3\,meV
(265~$\pm$~35\,K).\footnote{The error is estimated from the
difference of the least square fit and the Wannier function approach
for the TB model.} For the calculated range of physically relevant
$U_d$ values\footnote{Within the range of $U$ = 6...8.5\,eV we
obtain agreement with the experimentally reported exchange integrals
for a large number of edge- and corner-shared cuprate systems.} in
the LSDA+$U$ approach we obtain $J^{\mathrm{TH}}_{\mathrm{intra}}
\equiv J_1$~=~24~$\pm$~3\,meV (280~$\pm$~35\,K). The very good
agreement between $J_1^{\mathrm{AF}}$ and
$J^{\mathrm{TH}}_{\mathrm{intra}}$ leads us to the conclusion that
ferromagnetic contributions to the NN exchange in AgCuVO$_4$ are
basically negligible. The choice of $U_d$ is additionally justified
by the resulting gap size of 1.5...2\,eV consistent with the red
color of the sample and in agreement with measurements of the
absorption in the visible part of the electromagnetic spectrum at
significantly lower energy for AgCuVO$_4$ in comparison with
$\alpha$-AgVO$_3$ (reported gap of $\approx$ 2.3\,eV)\cite{konta} as
a reference.

A closer inspection of the effective exchange path by the calculated
Wannier functions (see Fig.~\ref{wannier}) shows that the
interaction does not appear from a coupling via the shared O$^{\mathrm{eq}}$
only, but also involves states of the [VO$_4$] tetrahedra. These
contributions are in line with the picture obtained from the orbital
resolved DOS (see Fig.~\ref{dos} and inset) with additional states
mixing into the antibonding $dp\sigma^*$ band.

It is well known that for ideal one-dimensional chains no magnetic
order appears.\cite{Mermin} In our compound, we calculated
inter-chain couplings $J^{\mathrm{TH}}_{\mathrm{inter}}$ of the
order of 1/65 with respect to the NN intra-chain exchange J$_1$ =
$J^{\mathrm{TH}}_{\mathrm{intra}}$. The calculated exchange values
allow the estimate of the ordering temperature $T_N$ according to
Eq.~\ref{neel}.\cite{schulz} Using $J_1$\,=\,24\,meV (280\,K) and an
effective $J^{\mathrm{TH}}_{\mathrm{inter}}$\,=\,0.25\,meV (3\,K) as
an average of the inter-chain couplings $J_1^{\mathrm{ic}}$ and
$J_3^{\mathrm{ic}}$ in the crystallographic $bc$ plane and the
inter-layer exchanges $J_1^{\mathrm{il}}$ and $J_2^{\mathrm{il}}$,
see Fig.~\ref{tbmodel}, we obtain a theoretical value
$T_N^{\mathrm{TH}}$\,=\,8.6(5)\,K for the N\'eel temperature.

\begin{figure}
\begin{center}
\resizebox{0.85\figwidth}{!}{
\includegraphics*[angle=0]{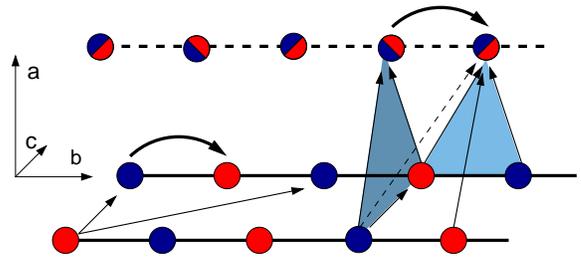}}
\caption{(Color online) Sketch of the magnetic model for AgCuVO$_4$.
The leading inter-chain exchange terms yield unfrustrated coupling
between the chains in the crystallographic $bc$ plane, but sizable
frustration between the layers (indicated by filled triangles).
This is illustrated by the different colors (red, blue) for
different spin directions according to the antiferromagnetic nature
of the relevant exchange couplings. }\label{magmodel}
\end{center}
\end{figure}

Compared to the experimentally observed $T_N$\,=\,2.5\,K this is a
clear overestimate by a factor of $\approx$ 3.5. A closer inspection
of the further magnetic interactions between chains belonging to
different 'magnetic layers' separated by a diagonal component along
the crystallographic $a$-direction ($J^{\mathrm{il}}$, cf.
Table~\ref{tandJ}) reveals the frustrating nature of the leading
inter-layer couplings $J_1^{\mathrm{il}}$ and $J_2^{\mathrm{il}}$
(compare Fig.~\ref{tbmodel} and see Fig.~\ref{magmodel}). Therefore,
we can assign the deviation of the calculated and observed $T_N$ to
the latter considerable inter-chain frustration. The situation is
comparable to the quasi 1D model compound Sr$_2$CuO$_3$, where a
similar suppression of the ordering temperature compared to the
calculation appears.\cite{rosner97} The strong influence of
frustration on the ordering temperature is especially pronounced for
the compound Sr$_2$Cu(PO$_4$)$_2$, where the leading inter-chain
couplings $J_\perp$ are fully frustrated.\cite{johannes2007}
Although the main couplings are comparable to AgCuVO$_4$, the
ordering temperature of Sr$_2$Cu(PO$_4$)$_2$ is smaller by a factor
of about 30. The strong quantum fluctuation in quasi 1D spin 1/2
compounds not only lead to a drastic reduction of the ordering
temperature, but also to a small ordered moment. Using the same
exchange parameters as for the calculation for $T_N$ we predict an
ordering moment of about 0.15~$\mu_B$ (Eq.~7 in Ref.
\onlinecite{schulz}).

In conclusion, our band structure calculations provide us with a
picture (see Fig.\ref{magmodel}) of quasi-1D NN Heisenberg chains
with small inter-chain couplings where the in-plane couplings
$J^{\mathrm{ic}}$ support AFM order, whereas sizable inter-layer
frustration $J^{\mathrm{il}}$ impedes AFM.

\section{Summary and Conclusions}

We have presented measurements of the static magnetic susceptibility
and the specific heat capacity, which evidence the pronounced one
dimensional magnetic properties of AgCuVO$_4$ related to structural
features and to the $d_{x^2-y^2}$ ground state character of the
Cu$^{2+}$ ion. Within the isotropic exchange model for a Heisenberg
system $J_{\mathrm{intra}} \approx$ 330\,K has been derived from the
experimental data. Furthermore, weak inter-chain coupling is evident
from specific heat capacity and ESR measurements leading to a
magnetically ordered state at $T_N$ = 2.5\,K. In order to gain
insight into the magnetic interactions on a microscopic basis we
performed full-potential electronic structure calculations.
Subsequently derived Heisenberg models based on Wannier functions
and LSDA+U calculations confirm the quasi 1D behavior. We found the
main exchange along the chain $J^{\mathrm{TH}}_{\mathrm{intra}}$ =
280~$\pm$ 35\,K in good agreement with the thermodynamic
measurements and several small inter-chain couplings leading to an
estimate for the antiferromagnetic ordering temperature
$T^{\mathrm{TH}}_N \sim$ 8.6\,K.  From the magnetic model (see
Fig.~\ref{magmodel}) the overestimate of $T^{\mathrm{TH}}_N$ by a
factor of 3.5 compared to the experimental observed ordering
temperature can be attributed to frustrated inter-chain couplings
along the crystallographic $a$ direction. All inter-chain couplings
are fairly weak and according to our experimental data and
calculations neither spiral order and subsequent multiferroic
behavior nor a Spin-Peierls transition is expected for AgCuVO$_4$.
Moreover, this compound with effective NN couplings of the order of
room temperature can be classified as 'in between' edge- and corner-
shared chains and might serve as a good candidate to study the
characteristic properties of a one-dimensional {\it S} = 1/2
Heisenberg system also by other complementary methods in temperature
ranges that are experimentally easier accessible. In particular,
single crystals are highly desirable for further investigation of
the thermal transport and expansion behavior as well as for studies
of the ordered magnetic moment and the magnon dispersion by neutron
diffraction.

\begin{acknowledgments}
This work was supported by the DFG thorough SFB 608. The authors would
like to thank Judith Jainski for the sample preparation, Oleg Janson,
J\"org Sichelschmidt and John A. Mydosh for valuable discussions and
U. Nitzsche for the use of the computational facilities at the IFW
Dresden.
\end{acknowledgments}




%



\end{document}